\documentclass[11pt]{article}

\usepackage{latexsym}
\usepackage{amsfonts}

\newtheorem{theorem}{Theorem}

                {\hfil\hfill$\square$\endtrivlist}

\begin{document}

\title{New electromagnetic conservation laws}

\author{G Bergqvist$^1$, I Eriksson$^1$, and J M M Senovilla$^2$ \\
${}^1$Matematiska institutionen, Link\"opings universitet, \\
      SE-581 83 Link\"oping, Sweden \\
${}^2$Departamento de F\'{\i}sica Te\'orica, Universidad del
Pa\'{\i}s Vasco, \\
Apartado 644, 48080 Bilbao, Spain \\
gober@mai.liu.se, ineri@mai.liu.se, wtpmasej@lg.ehu.es}

\maketitle

\begin{abstract}
The Chevreton superenergy tensor was introduced in 1964 as a
counterpart, for electromagnetic fields, of the well-known
Bel-Robinson tensor of the gravitational field.
We here prove the unnoticed facts that, in the absence of
electromagnetic currents,
Chevreton's tensor (i) is completely symmetric, and (ii) has a
trace-free divergence if Einstein-Maxwell equations hold.
It follows that the trace of the Chevreton tensor
is a rank-2, symmetric, trace-free, {\em conserved} tensor, which is
different from the energy-momentum tensor, and nonetheless
can be constructed for any test Maxwell field,
or any Einstein-Maxwell spacetime.

\end{abstract}


\newpage
The Bel-Robinson ``superenergy'' tensor \cite{Bel1,Bel3}
is today a well-known tool in General
Relativity. Despite the lack of a conclusive physical meaning, it has
been proved as very valuable in many mathematical developments and theoretical
applications, see e.g.\cite{PR,seno2000} and references therein.
The analogy of many of its properties with those of the
energy-momentum tensor of electromagnetic fields is intriguing and
certainly suggestive, something which has led many authors to look
for similar superenergy tensors of fields other than the graviational
one (e.g \cite{chevreton,PR,seno2000,Tey} and references therein).
Perhaps the first such attempt appears in the work by Chevreton
\cite{chevreton}, who introduced a 4-index tensor, similar to the
Bel-Robinson one, for the Maxwell field. Chevreton's tensor is
quadratic in the derivatives of the electromagnetic 2-form, and
divergence-free in the absence of gravitational field, that is, in
Special Relativity. This last property does not hold in the presence
of curvature \cite{chevreton,seno2000}, possibly leading to the exchange of
superenergy between the fields, see e.g. \cite{seno2000,rujora}.

The purpose of this letter is to prove two apparently unnoticed
properties of Chevreton's tensor. In the absence of electromagnetic
sources we prove first that the Chevreton tensor is completely
symmetric, and then that its trace is divergence-free (or, equivalently,
its divergence is trace-free) if either Einstein's vacuum or 
Einstein-Maxwell's field equations hold.

We will follow the spinor and abstract index notations as
defined in \cite{PR} and use the signature $(+ - - -)$ (observe that this
is opposite to \cite{S1,seno2000}). The metric tensor will be
denoted by $g_{ab}$.
The definition of the Chevreton superenergy tensor can today, as
is also the case for the Maxwell, Bel \cite{Bel2} and Bel-Robinson tensors, be
seen as a construction that comes from the general definition of
superenergy tensors in Lorentzian manifolds of arbitrary dimension.
This general definition provides an even rank tensor $T_{ab\dots}\{
A_{\dots}\}$ starting from any arbitrary tensor $A_{ab\dots}$, the
former being quadratic on the latter and called the basic superenergy tensor of
$A_{ab\dots}$. The definition was originally presented in \cite{S1}
and studied and much developed in \cite{seno2000}. Following this
paper, a tensor $A_{abc}$ with the antisymmetry
$A_{abc}=A_{a[bc]}$ is called a double (1,2)-form and its basic superenergy
tensor $T_{abcd}\{ A_{[1],[2]}\}$ can be adequately expressed as
\cite{seno2000}
\begin{eqnarray}\label{eq:se12g}
T_{abcd}\{ A_{[1],[2]}\}=-A_{acf}A_{bd}{}^{f}-A_{adf}A_{bc}{}^{f}
+ g_{ab}A_{ecf}A^e{}_{d}{}^{f}
+ {1\over 2}g_{cd}A_{aef}A_{b}{}^{ef}
-{1\over 4}g_{ab}g_{cd}A_{efg}A^{efg}
\end{eqnarray}
which is explicitly independent of the dimension of the spacetime and
satisfies $T_{abcd}\{ A_{[1],[2]}\}=T_{(ab)(cd)}\{ A_{[1],[2]}\}$.

Let $F_{ab}=-F_{ba}$ be a 2-form and consider the (1,2)-form
$A_{abc}=\nabla_a F_{bc}$. Its basic superenergy tensor
$T_{abcd}\{\nabla_{[1]}F_{[2]}\}$ is given then by
(\ref{eq:se12g}). We will denote it by $E_{abcd}$ in what follows.
If $F_{ab}$ is a Maxwell field, Chevreton's superenergy
tensor of the electromagnetic field is defined by \cite{chevreton,WZ}
\begin{equation}\label{eq:chev}
H_{abcd}={1\over 2}(T_{abcd}\{\nabla_{[1]}F_{[2]}\}+
T_{cdab}\{\nabla_{[1]}F_{[2]}\})\equiv {1\over 2}(E_{abcd}+E_{cdab})
\end{equation}
so that it has the obvious symmetries
$$
H_{abcd}=H_{(ab)(cd)}=H_{cdab}.
$$
It also satisfies the following positive-definite property ---called the
dominant property \cite{bergqvist01}---:
$$
H_{abcd}u^av^bw^cz^d\ge 0
$$
for all future-directed causal vectors $u^a$, $v^a$, $w^a$
and $z^a$. This follows from (\ref{eq:chev}) and the fact that {\it any}
superenergy tensor has the dominant property \cite{bergqvist99,
seno2000}, as well as any linear combination of them with non-negative
coefficients \cite{seno2000,bergqvist01}. As mentioned before,
in flat spacetime $E_{abcd}$ is divergence-free.
Thus, the analogy with the Bel-Robinson tensor is quite
stimulating. We are going to make this analogy even stronger in
the remaining of this letter.

   From now on, we assume that the spacetime is 4-dimensional.\footnote{In this
case, an analysis of the Chevreton tensor of null electromagnetic fields
can be found in \cite{WZ}.} Then, as is well-known, the Bel-Robinson tensor is
completely symmetric and traceless. We are going to obtain the corresponding
properties for the Chevreton tensor. Concerning traces, we first of
all have \cite{seno2000}
\begin{equation}
E_{ab}{}^c{}_{c}=0, \hspace{3mm} E^c{}_{c}{}_{ab}\neq 0,
\hspace{3mm} E^c{}_{abc}\neq 0 \label{pretraces}
\end{equation}
so that in general the traces of $H_{abcd}$ are non-vanishing.
Nevertheless, they satisfy some interesting properties. To prove
them, and to obtain the result on the index symmetry, we find it
easier to resort to the spinor formalism \cite{PR}. We are going to prove
that $H_{abcd}$ is completely symmetric if $F_{ab}$ satisfies
the source-free Maxwell equations.

The general spinor form of superenergy tensors in dimension 4
was obtained in \cite{bergqvist99}. For the (1,2)-form
$A_{abc}=A_{AA'BB'CC'}$ this is
$$
T_{abcd}\{ A_{[1],[2]}\}=E_{abcd}={1\over 2}(A_{AB'CE'D}{}^{E'}
\bar A_{BA'EC'}{}^{E}{}_{B'}+A_{BA'CE'D}{}^{E'}
\bar A_{AB'EC'}{}^{E}{}_{B'})
$$
so that writing $F_{ab}$ in spinor form
$$
F_{ab}=\varphi_{AB}\bar\epsilon_{A'B'}+\bar\varphi_{A'B'}
\epsilon_{AB}
$$
where $\varphi_{AB}=\varphi_{(AB)}={1\over 2}F_{AE'B}{}^{E'}$
we get for $A_{abc}=\nabla_a F_{bc}$ that
$$
A_{AB'CE'D}{}^{E'}=2\nabla_{AB'}\varphi_{CD}
$$
hence its basic superenergy tensor becomes
$$
E_{abcd}=2(\nabla_{AB'}\varphi_{CD}
\nabla_{BA'}\bar\varphi_{C'D'}+\nabla_{BA'}\varphi_{CD}
\nabla_{AB'}\bar\varphi_{C'D'}).
$$
The general spinor form of the Chevreton tensor is therefore
\begin{eqnarray}\label{eq:spinchev}
H_{abcd}&=\nabla_{AB'}\varphi_{CD}
\nabla_{BA'}\bar\varphi_{C'D'}+\nabla_{BA'}\varphi_{CD}
\nabla_{AB'}\bar\varphi_{C'D'} \nonumber \\
&+\nabla_{CD'}\varphi_{AB}
\nabla_{DC'}\bar\varphi_{A'B'}+\nabla_{DC'}\varphi_{AB}
\nabla_{CD'}\bar\varphi_{A'B'}\, .
\end{eqnarray}
In the absence of electromagnetic sources, the Maxwell equations
$$
\nabla^a F_{ab}=0, \qquad \nabla_{[a}F_{bc]}=0
$$
are equivalent to \cite{PR}
\begin{equation}
\nabla^{AA'}\varphi_{AB}=0 \label{Maxw}
\end{equation}
which implies that
\begin{equation}
\nabla_{B'A}\varphi_{CD}=\nabla_{B'(A}\varphi_{CD)} \label{sym1}
\end{equation}
so the Chevreton tensor becomes in this case
\begin{eqnarray*}
H_{abcd}=\nabla_{B'(A}\varphi_{CD)}
\nabla_{B(A'}\bar\varphi_{C'D')}+\nabla_{A'(B}\varphi_{CD)}
\nabla_{A(B'}\bar\varphi_{C'D')} \hspace{17mm} \nonumber \\
+\nabla_{D'(C}\varphi_{AB)}
\nabla_{D(C'}\bar\varphi_{A'B')}+\nabla_{C'(D}\varphi_{AB)}
\nabla_{C(D'}\bar\varphi_{A'B')} =H_{(abcd)}
\end{eqnarray*}  
Hence we have proved the following new result:
\begin{theorem}\label{th:symm}
        In the absence of electromagnetic sources,
				the Chevreton tensor is completely
symmetric in four dimensions: $H_{abcd}=H_{(abcd)}$.
\end{theorem}
Observe that this theorem is valid in arbitrary spacetimes,
independently of Einstein's field equations: only the Maxwell
equations have been used. We want to remark that Theorem \ref{th:symm}
can also be derived as an application of the
anti-symmetrization methods developed by Edgar and H{\"o}glund
\cite{eh2002}, which in fact inspired us.
Their methods would lead to an alternative tensorial proof of the theorem.

As a corollary of the above theorem, we obtain that the two non-zero
traces of $E_{abcd}$ are related by means of
\begin{equation}
\frac{1}{2}E^c{}_{c}{}_{ab}=E^c{}_{abc}=E^c{}_{(ab)c}=
H^c{}_{c}{}_{ab} \equiv H_{ab}\label{traces}
\end{equation}
as follows directly from (\ref{eq:chev}) and the complete symmetry of
$H_{abcd}$. Moreover, due to the first expression in
(\ref{pretraces}) we have also proved that the complete trace of
$H_{abcd}$ (or $E_{abcd}$) vanishes identically in the absence of
sources, in other words that the tensor $H_{ab}$ is trace-free:
$$
H^a{}_{a}=0.
$$

We now go on to express the divergence of the Chevreton tensor
and show that it is trace-free in the absence of sources. Equivalently,
this can be expressed as saying that the trace $H_{ab}$ of the Chevreton
tensor, as defined in (\ref{traces}), is divergence-free.
Note however that the trace of a
superenergy tensor does not need to have the dominant property.

In general there are two divergences of $E_{abcd}$ which are
$\nabla^a E_{abcd}$ and $\nabla^c E_{abcd}$. Explicit expressions for
them, rather lengthy in general, are given in \cite{seno2000} in terms of
the curvature tensor. For the completely symmetric Chevreton tensor $H_{abcd}$
we get of course only one independent divergence. Here we use the
spinor form (\ref{eq:spinchev}) directly to get
\begin{eqnarray}\label{eq:divchev}
\nabla ^a H_{abcd} &
= & \nabla ^{AA'} \nabla _{AB'} \varphi _{CD} \nabla _{A'B} \bar \varphi
_{C'D'}  + \nabla _{AB'} \varphi _{CD} \nabla ^{AA'} \nabla _{A'B} \bar
\varphi _{C'D'}   \nonumber\\
&   & +\nabla ^{AA'} \nabla _{BA'} \varphi
_{CD} \nabla _{B'A} \bar \varphi _{C'D'}  + \nabla _{BA'} \varphi _{CD}
\nabla ^{AA'} \nabla _{B'A} \bar \varphi _{C'D'}   \nonumber\\
     &   & +\nabla ^{AA'} \nabla _{CD'} \varphi _{AB} \nabla _{C'D} \bar
\varphi _{A'B'}  + \nabla _{CD'} \varphi _{AB} \nabla ^{AA'} \nabla
_{C'D} \bar \varphi _{A'B'}   \nonumber\\
      &   & +\nabla ^{AA'} \nabla _{DC'} \varphi _{AB}
\nabla _{D'C} \bar \varphi _{A'B'}  + \nabla _{DC'} \varphi _{AB}
\nabla ^{AA'} \nabla _{D'C} \bar \varphi _{A'B'}
\end{eqnarray}
To study this expression we use commutators and note that there are
essentially two types of terms, represented by $\nabla ^{AA'} \nabla
_{AB'}\varphi _{CD}$ and $\nabla ^{AA'} \nabla _{BA'} \varphi_{CD} $.
Applying (\ref{sym1}) when needed we get, with the notation of \cite{PR}
and use of (\ref{Maxw}),
\begin{eqnarray}\label{kom1}
& \nabla^{AA'} \nabla _{AB'} \varphi _{CD}=
\nabla^{AA'} \nabla _{CB'} \varphi _{AD} \nonumber\\
& =\nabla _{CB'} \nabla ^{AA'} \varphi _{AD} +
(\bar\varepsilon^{A'}{}_{B'}\Box^{A}{}_{C}+
\varepsilon^{A}{}_{C}\Box^{A'}{}_{B'})\varphi _{AD} \nonumber\\
& =\bar\varepsilon_{B'}{}^{A'}(
X^E{}_{CE}{}^{F} \varphi_{FD} + X^E{}_{CD}{}^{F} \varphi_{EF})+
\Phi^{A'}{}_{B'C}{}^{F} \varphi_{FD} + \Phi^{A'}{}_{B'D}{}^{F}
\varphi_{CF}
\end{eqnarray}
and, also using the contraction of (\ref{kom1}),
\begin{eqnarray}\label{kom2}
& \nabla_A{}^{A'} \nabla _{BA'} \varphi_{CD}=
(\nabla_{(A}{}^{A'} \nabla _{B)A'} + \frac{1}{2}
\varepsilon_{AB}\nabla_E{}^{A'}\nabla^E{}_{A'})\varphi _{CD} \nonumber\\
& =-(\Box_{AB} + \frac{1}{2}\varepsilon_{AB}
\nabla^{e}\nabla_e) \varphi _{CD}  \nonumber\\
& =X_{ABC}{}^{F}\varphi_{FD} +X_{ABD}{}^{F} \varphi_{CF}+
\varepsilon_{BA}(
X^E{}_{CE}{}^{F} \varphi_{FD} + X^E{}_{CD}{}^{F} \varphi_{EF}).
\end{eqnarray}
Here $\Phi_{ABA'B'}$ is the trace-free Ricci spinor and
$X_{ABCD}=\Psi_{ABCD}+\Lambda(\varepsilon_{AC}\varepsilon_{BD}+
\varepsilon_{AD}\varepsilon_{BC})$ with $\Psi_{ABCD}$ being the
Weyl spinor and $24\Lambda=R$ the scalar curvature.
Substituting (\ref{kom1}) and (\ref{kom2}) into (\ref{eq:divchev})
and imposing part of the Einstein-Maxwell equations (namely,
$\Phi_{ab}=2\varphi_{AB}\bar\varphi_{A'B'}$ but keeping $\Lambda$
arbitrary) we find after some simplifications and use of (\ref{sym1})
and Theorem \ref{th:symm} the following simple expression
\begin{eqnarray}
      \nabla ^a H_{abcd}
=2\Psi^{EF}{}_{(BC} \varphi _{D)F}
\nabla_{E(B'} \bar \varphi_{C'D')} +
4 \varphi^{FE} \Psi_{FE(BC}\nabla _{D)(B'} \bar \varphi_{C'D')}
-18\Lambda\varphi_{(BC}\nabla_{D)(B'}\bar\varphi_{C'D')}\nonumber\\
+2\bar\Psi^{E'F'}{}_{(B'C'} \bar\varphi _{D')F'}
\nabla_{E'(B} \varphi _{CD)} +
4\bar\varphi^{F'E'}\bar\Psi_{F'E'(B'C'}\nabla _{D')(B}\varphi _{CD)}
-18\Lambda\bar\varphi_{(B'C'}\nabla_{D')(B}\varphi_{CD)}.
\label{divche}
\end{eqnarray}
Obviously, this expression holds in Einstein spaces too
($\Phi_{ab}=0$ with any $\Lambda$).
Since (\ref{divche}) is completely symmetric with respect to spinor
indices it is trace-free, that is to say,
$$
\nabla ^a H_{ab}=0.
$$
Thus, we have the new result
\begin{theorem}\label{th:div}
      The trace of the Chevreton tensor is divergence-free in the absence of
electromagentic sources in four dimensions.
\end{theorem}
We stress the fact that this result holds (i) for arbitrary
Einstein-Maxwell spacetimes with a possible cosmological constant
$\Lambda$ as well as (ii) for any Maxwell test fields
in Einstein spaces, including proper vacuum.
Therefore, we have constructed a symmetric, trace-free and
divergence-free tensor $H_{ab}$ associated to any source-free electromagnetic
field. This tensor is quadratic in the derivatives of $F_{ab}$ and
therefore it is not related to the energy-momentum tensor (nor to the
so-called zilch tensor \cite{lip,kib} which is conserved in Special
Relativity, see also \cite{AP,FN}).
Note that by (\ref{eq:spinchev}) and (\ref{traces}) we have the simple spinor
form of $H_{ab}$
$$
H_{ab}=-2\nabla_c\varphi_{AB}\nabla^c\bar\varphi_{A'B'}
$$
which can be used to derive the above theorem directly. Several
convenient tensor forms are
\begin{eqnarray*}
H_{ab}=\nabla_c F_{ad}\nabla^c F_b{}^d-
\frac{1}{4}g_{ab}\nabla_c F_{de}\nabla^c F^{de}\\
=\frac{1}{2}\left(\nabla_c F_{ad}\nabla^c F_b{}^d+
\nabla_c \stackrel{*}{F}_{ad}\nabla^c \stackrel{*}{F}_b{}^d\right)\\
=2\nabla_c\ {}^+F_{ad}\ \nabla^c\ {}^-F_b{}^d \hspace{1cm}
\end{eqnarray*}
where $\stackrel{*}{F}_{ab}$ is the Hodge dual of $F_{ab}$ and
${}^{\mp}F_{ab}\equiv \frac{1}{2}(F_{ab}\pm i \stackrel{*}{F}_{ab})$. 
We wish to remark that the proof of the divergence-free property of
$H_{ab}$ is far from obvious when using these tensor expressions.

One can construct conserved currents associated to
$H_{ab}$ if Killing or conformal Killing vectors are present.
Explicitly, they are given by
$$
j^a(\xi) \equiv H^a{}_{b}\xi^b \hspace{2mm} \Longrightarrow 
\hspace{2mm} \nabla_{a}j^a=0
$$
where $\xi^b$ is any (conformal) Killing vector. These currents, as
well as $H_{ab}$, are invariant against duality
rotations\footnote{The fitting of this result in
Special Relativity with those of \cite{AP,FN} is unclear.}
${}^{\pm}F_{ab}\rightarrow e^{\pm i\theta}\ {}^{\pm}F_{ab}$ (with
constant $\theta$)
as follows from the fact that the basic superenergy tensor of any tensor
coincides with that of any of its duals \cite{seno2000}, or directly
from the explicit expressions of $H_{ab}$.
Observe on the other hand that, generically, the dominant property
(the dominant energy  condition), i.e. that $H_{ab}u^av^b\ge 0$
for all future-directed causal vectors $u^a$ and $v^a$, need not be
satisfied.

\subsection*{Acknowledgements}
We thank Brian Edgar for discussions and valuable
comments.

\end{document}